\documentclass[preprint,prc,nofootinbib,superscriptaddress,a4]{revtex4}
\usepackage{color}
\usepackage{amssymb}
\usepackage{ulem}
\usepackage{amsmath}
\usepackage{graphicx}

\usepackage{xcolor}
\def\be{\begin{equation}} \def\ee{\end{equation}} \def\bea{\begin{eqnarray}}
\def\eea{\end{eqnarray}} \def\nnb{\nonumber}

\begin{document}
\title{
Analysis of elastic $\alpha$-$^{12}$C scattering with global optimization
in the cluster effective field theory
}
\author{Myeong-Hwan Mun}
\affiliation{Department of Physics, Kyungpook National University, Daegu 41566, Korea}
\author{Jubin Park}
\email{honolov@ssu.ac.kr}
\affiliation{Department of Physics and Origin of Matter and Evolution of Galaxies Institute,
Soongsil University, Seoul 06978, Korea}
\author{Chang Ho Hyun}
\affiliation{Department of Physics Education, Daegu University, Gyeongsan 38453, Korea}
\author{Shung-Ichi Ando}
\affiliation{Department of Physics Education, Daegu University, Gyeongsan 38453, Korea}
\affiliation{Department of Display and Semiconductor Engineering and Research Center for Nano-Bio Science, Sunmoon University, Asan 41439, Korea}

\date{\today}

\begin{abstract}
We analyze the elastic $\alpha$-$^{12}$C scattering including
the contribution of resonance states below 
the $p$-$^{15}$N breakup threshold energy.
We use the cluster effective field theory in which scattering
amplitude is expanded in terms of the effective range expansion
parameters for the angular momentum states from $l=0$ to $l=6$.
The amplitude contains 37 parameters, which are determined by fitting to
11~392 differential cross section data points of the elastic $\alpha$-$^{12}$C scattering.
To optimize the fitting process, 
we implement the differential evolution (DE) algorithm,
which performs a global search over the high-dimensional parameter space and
consistently converges to the same minimum $\chi^{2}$ value across independent runs,
suggesting proximity to the global minimum within the explored domain.
In parallel, the Markov chain Monte Carlo (MCMC) method is used to crosscheck the DE results and to estimate the parameter uncertainties.
The best fit yields $\chi^{2}/N\!\simeq\!6.2$ for the elastic scattering data.
Using the determined 37 parameters, we calculate the differential
cross sections and the phase shifts of the elastic $\alpha$-$^{12}$C
scattering and compare the results with experimental data and
those of an $R$-matrix analysis.
Our result of the cross section agrees with the experimental data
as accurately as an $R$-matrix analysis.
The results demonstrate that the cluster effective field theory,
combined with global optimization and uncertainty quantification based on DE-MCMC methods,
provides a reliable and systematic framework for applications to low energy phenomena relevant to stellar evolution and nucleosynthesis.
\end{abstract}

\maketitle

\section{Introduction}
Effective field theory (EFT) in nuclear physics theory was initiated with the purpose
to describe the strong interaction between two nucleons in free space systematically
and perturbatively~\cite{w-pa79,w-plb90}. 
In the EFT energy and momentum in a system or process
under consideration are limited to an upper limit, which roughly divides the region
above which EFT breaks down.
This upper scale is combined with the counting rules that determine the order of 
reaction amplitudes 
for a specific process in the power of energy or momentum 
divided by the upper scale, and it becomes feasible to describe 
two-nucleon phenomenology at low energies on a field theoretic ground~\cite{hkvk-rmp20}. 

Depending on the magnitude of energy and momentum under consideration,
scale parameter and counting rules are adjusted to optimize the simplicity
and efficiency of the theory.
For instance, if one is considering processes in the big bang nucleosynthesis era, relevant energy is below 1 MeV.
At this energy scale, even the pion becomes a heavy degree of freedom.
Pions are integrated out, and interactions can be described in the contact form.
This theory is called pionless EFT~\cite{crs-npa99,k-npb97}. 
It showed accurate description of two-nucleon systems \cite{prc2005, prc2007, plb2008} and
interactions with external electromagnetic and weak probes \cite{prc2006, prc2010, prc2011,
prc2013, jkps2016, prc2017, prc2020}.
The idea of pionless EFT is extended to the systematic expansion of nuclear 
energy density functional to describe the infinite nuclear matter and finite 
nuclei, and the attempt resulted in the birth of KIDS (Korea-IBS-Daegu-SKKU)
density functional \cite{kids1, kids2, kids3, kids4, kids5}.
Works with KIDS functional have been showing successful applications to diverse 
nuclear many-body problems such as nuclear structure \cite{st1, st2, st3, st4, st5}, 
lepton-nucleus scattering \cite{ln1, ln2, ln3, ln4, ln5} and 
nuclear astrophysics \cite{astro1, astro2, astro3, astro4, astro5}.

Works applying the EFT and its extensions, including the KIDS functional, to astrophysical phenomena become more and more popular nowadays.
In the stellar evolution and synthesis of heavy elements, the radiative capture of
$\alpha$ particle by $^{12}$C, $^{12}$C($\alpha, \gamma$)$^{16}$O plays an essential role.
The process primarily happens in the core helium burning stage of the stellar evolution.
Temperature at this stage is well below 1 MeV, a scale much smaller than the pion mass,
so interactions that drive the capture process
can be described in terms of the pionless theory by treating $\alpha$,
$^{12}$C, and $^{16}$O as point particles.
This theory is named a cluster EFT.
Application of the cluster EFT to elastic $\alpha$-$^{12}$C scattering showed that
by adjusting the effective range expansion (ERE) parameters in the theory  to the
phase shifts of elastic $\alpha$-$^{12}$C scattering, one can reproduce 
the experimental data accurately in the angular momentum states from $l=0$
to $l=6$ \cite{sa-epja16, sa-prc18, sa-prc20, sa-prc22, sa-prc23}.
The success proves that the cluster EFT could be a candidate for a theory
applicable to the study of stellar evolution and nucleosynthesis of light nuclei.

While the $\alpha$-$^{12}$C capture process is so important in nuclear astrophysics,
there is a critical difficulty in understanding the process.
The capture process happens
most actively at Gamow energy $E_G\simeq 0.3$ MeV within stars.
Because the Coulomb barrier is much higher than $E_G$, it is extremely difficult to
measure the capture cross section (or equivalently $S$ factor) at $E_G$ experimentally.
For precise understanding of stellar evolution and synthesis of elements,
it is inevitable to rely on theoretical estimation of the capture cross section.
However, the status of the theory is still ambiguous that recent estimations suffer from
uncertainties about 20--30~\%, e.g., $S_{E1}(E_G) = 80\pm18$ keV b by Tang {\it et al.} \cite{tang} 
and $100 \pm 28$ keV b by Oulebsir {\it et al.} \cite{oulebsir}.
To reduce the gap, a new method may have to be attempted and advanced constraints on the
uncertainty are demanded.

In this work, we adopt the cluster EFT to analyze the elastic scattering of $\alpha$ particle
and $^{12}$C.
Undetermined constants of the theory corresponding to ERE parameters for the angular momentum from $l=0$ to $l=6$ are fitted to the elastic $\alpha$-$^{12}$C scattering 
cross section data in the energy range 2.6--6.7 MeV and 32 angle values
in the range 24.0$^\circ$--165.9$^\circ$~\cite{tetal-prc09}. 
Total number of data amounts to 11~392.
To efficiently handle the large number of data and to obtain optimized values of the fitted parameters, we employ two complementary numerical techniques.
First, we adopt the differential evolution (DE) algorithm to determine the ERE parameters
of the theory.
It is known that DE is a method optimized to find global minimum over a large number
of parameter space.
Once the parameters are determined from the DE calculation,
we use the values of DE as a starting point of the Markov chain Monte Carlo (MCMC) calculation.
There are two purposes to run the MCMC algorithm: the first is a crosscheck of the 
results obtained by the DE calculation, and the second is to estimate the 
uncertainties of the parameters.
Estimation of the uncertainty is particularly important in the extrapolation of the theory
to astrophysical energy scales.
It is claimed that to have a quantitative description of the helium burning process,
the capture cross section at $E_G$ should be controlled with uncertainties less than 10~\% \cite{deboer}.

As a result of the fitting with DE algorithm,
we obtain $\chi^2/N \simeq 6.23$ for the $N=11~392$ differential cross section data.
Plugging the DE results of parameters in the starting values of MCMC fitting,
we obtain slightly improved accuracy $\chi^2/N\simeq 6.18$.
We calculate the 11~392 differential cross section points with the 37 parameters determined from the MCMC fitting and compare the results with experimental data and $R$-matrix
calculation.
Cluster EFT reproduces the experimental cross section data as accurately as the
$R$-matrix theory.\footnote{
For a visual comparison with $R$-matrix, see Fig.~6 and discussions of 
Ref.~\cite{tetal-prc09}.}
To further test the theory, phase shifts are calculated from $l=0$ to $l=6$
and the results are compared with two $R$-matrix calculations \cite{tetal-prc09, petal-npa87}.
Most of the phase shifts from $l=0$ to $l=4$ agree well between cluster EFT and 
$R$-matrix theory.
However, for $l=5$ and $l=6$, results of cluster EFT are consistent with the
$R$-matrix results in Ref.~\cite{petal-npa87}, but there are substantial discrepancies
with Ref.~\cite{tetal-prc09}.
Finally asymptotic normalization constants (ANCs) of $^{16}$O are calculated
for the $0^+_1$, $0^+_2$, $1^-_1$ and $3^-_1$ states and the result is compared
with those of previous EFT in which parameters are adjusted to the phase
shift data.
ANCs of $0^+_1$ and $1^-_1$ in this work are consistent with those in the 
previous EFT calculation.
On the other hand, the value of $0^+_2$ is higher in the new result,
and it is suppressed significantly in the $3^-_1$ state.

This study establishes a new data-driven framework that integrates the cluster EFT with DE-MCMC optimization.
Unlike conventional $R$-matrix or manually tuned analyses, the present approach performs 
an automated exploration of the entire high-dimensional parameter space, 
thereby avoiding dependence on initial guesses 
and preventing the fit from being trapped in local minima.
The sequential use of the DE and MCMC algorithms 
ensures statistically robust convergence 
and provides a consistent quantification of parameter correlations and uncertainties.
This methodology offers a reproducible and extensible strategy 
for determining the effective range parameters directly from experimental data, 
and it lays the groundwork for reliable extrapolation of the theory 
to astrophysical energy scales.

The paper is organized in the following order.
In Sec. II, we describe construction of the cluster EFT and how it is applied to the elastic
$\alpha$-$^{12}$C scattering.
In Sec. III, we introduce the method to apply DE-MCMC optimization to the analysis of elastic
$\alpha$-$^{12}$C scattering and present the fitting results and the associated uncertainties.
In Sec. IV, as applications of the fitting result, 
we present the result of the calculation of differential cross sections
and phase shifts in the elastic $\alpha$-$^{12}$C scattering, and ANCs
of the bound states of $^{16}$O.
Section V is dedicated to the summary and concluding remarks of the work.
Appendix A shows the detailed relations in the transformation between
laboratory frame and center-of-mass frame and
Appendix B explains in detail how to fix the effective range parameters.
Appendix C summarizes the numerical values of 37 effective range parameters
determined from DE and MCMC calculations.

\section{Elastic $\alpha$-$^{12}$C scattering in cluster EFT}

In this section, we review the formalism of the differential cross section, $S$ matrices, 
and scattering amplitudes of elastic $\alpha$-$^{12}$C scattering at low energies within the framework of the cluster EFT.
The differential cross section of the elastic $\alpha$-$^{12}$C scattering 
in the center-of-mass frame is given as~\cite{petal-npa87}
\bea
\sigma(\theta) = \frac{1}{p^2}\left|
-\frac{\eta}{2\sin^2\theta/2}e^{2i\sigma_0}e^{-i\eta\ln\sin^2\theta/2}
+ \frac{1}{2i}\sum_{l=0}^\infty (2l+1) e^{2i\sigma_l}\left(
S_l
-1\right)
P_l(\cos\theta)
\right|^2\,,
\eea
where $\theta$ is the scattering angle, 
$p$ is the magnitude of relative momentum, $p=\sqrt{2\mu E}$, where
$\mu$ is the reduced mass of $\alpha$ and $^{12}$C, and $E$ is the energy 
of $\alpha$-$^{12}$C system in the center-of-mass frame.
$\eta$ is the Sommerfeld factor, $\eta=\kappa/p$: 
$\kappa$ is the inverse of the Bohr radius, $\kappa=Z_\alpha Z_C\alpha_E \mu$.
$Z_\alpha$ and $Z_C$ are the numbers of protons in $\alpha$ and $^{12}$C,
and $\alpha_E$ is the fine structure constant, $\alpha_E=1/137.036$. 
$\sigma_l$ are the Coulomb phase shifts for $l$th partial waves, 
\bea
e^{2i\sigma_l} &=& \frac{\Gamma(l+1+i\eta)}{\Gamma(l+1-i\eta)},
\eea 
where $\Gamma(z)$ is the Gamma function. 
$S_l$ are the $S$ matrices of the elastic $\alpha$-$^{12}$C scattering
for $l$th partial wave states. $P_l(x)$ are the Legendre polynomial functions.
The differential cross sections are measured in the laboratory frame.
The formulas for the conversion of the cross section and scattering angle
between the laboratory frame and the center-of-mass frame are
given in Appendix A. 

The $S$ matrices of the elastic $\alpha$-$^{12}$C scattering for 
$l$th partial waves are given in terms of the phase shifts, $\delta_l$,
as
\bea
S_l &=& e^{2i\delta_l}=1+2ip\tilde{A}_l\,,
\eea
where $\tilde{A}_l$ is the scattering amplitude for $l$th partial waves. 
We now assume that the phase shifts can be decomposed, e.g., as~\cite{sa-prc22,sa-prc23}
\bea
\delta_l &=& \delta_l^{(bs)} + \delta_l^{(rs1)} + \delta_l^{(rs2)} + \cdots\,,
\label{eq:delta_l}
\eea
where $\delta_l^{(bs)}$ is the phase shift generated from the scattering amplitude of 
a bound state of $^{16}$O, and $\delta_l^{(rs1)}$ and $\delta_l^{(rs2)}$ are the phase shifts 
generated from the first and second resonant states of $^{16}$O for the $l$th angular momenta.  
Each part of the phase shift, $\delta_l$, is calculated by using a relation
\bea
e^{2i\delta_l^{(ch)}} &=& 1 +2ip\tilde{A}_l^{(ch)}\,,
\eea
where $(ch)annel=(bs), (rs1), (rs2), \cdots$, and $\tilde{A}_l^{(ch)}$ are the scattering amplitudes, 
which are calculated from the effective Lagrangian in theory. 
We note that the decomposition of the phase shift assumed in Eq.~(\ref{eq:delta_l})
is crucial to sum up the amplitudes, $\tilde{A}_l^{(ch)}$, 
preserving the unitary condition of the $S$ matrices. 

\begin{figure}[t]
\centering
\includegraphics[width=0.8\textwidth]{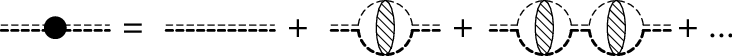}
\caption{
Feynman diagrams for dressed $^{16}$O propagators.
A thin dashed line and a thick dashed line denote the propagation of $\alpha$ and $^{12}$C, respectively.
A thin-thick double-dashed line with or without a filled circle denotes a dressed or bare $^{16}$O propagators.
A shaded oval denotes the Coulomb Green's function. 
}
\label{fig:propagators}
\end{figure}

\begin{figure}[t]
\centering
\includegraphics[width=0.23\textwidth]{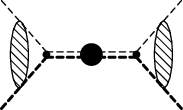}
\caption{
Feynman diagram for the scattering amplitudes of 
elastic $\alpha$-$^{12}$C scattering in cluster EFT.
A shaded oval denotes the initial or final state Coulomb wavefunction.
See the caption of Fig.~\ref{fig:propagators} as well. 
}
\label{fig:scattereing_amplitudes}
\end{figure}

In Figs.~\ref{fig:propagators} and \ref{fig:scattereing_amplitudes}, 
the Feynman diagrams of the dressed $^{16}$O propagators for $l$th partial waves and elastic 
$\alpha$-$^{12}$C scattering are displayed, respectively.\footnote{
We note that the counting rules of the scattering amplitudes are not readable 
through the diagrams in Figs. \ref{fig:propagators} and  \ref{fig:scattereing_amplitudes}. 
Refer to Ref.~\cite{sa-prc18} for the details of the modified counting rules employed 
in this work. 
}
The propagators and vertex functions are extracted from the effective Lagrangian 
in Eq.~(6) in Ref.~\cite{sa-prc23}, and the scattering amplitudes have been obtained as follows. 

The scattering amplitudes for the subthreshold bound $0_2^+$, $1_1^-$, $2_1^+$, $3_1^-$ 
($l_{i\textrm{th}}^\pi$) states of
$^{16}$O were obtained as~\cite{sa-prc22}
\bea
\tilde{A}_l^{(bs)} &=& \frac{C_\eta^2 W_l(p)}{K_l(p)-2\kappa H_l(p)}\,,
\eea
where the functions $C_\eta^2W_l(p)$ in the numerator of the amplitudes are calculated from the
initial and final state Coulomb wavefunctions in Fig.~\ref{fig:scattereing_amplitudes}, and 
\bea
C_\eta^2 &=& \frac{2\pi\eta}{\exp(2\pi\eta)-1}\,,
\\
W_l(p) &=& \left(
\frac{\kappa^2}{l^2} + p^2
\right)W_{l-1}(p)\,, \ \ \ 
W_0(p) = 1\,.
\eea
The functions $-2\kappa H_l(p)$ in the denominator of the amplitudes are the self-energy terms
calculated from the bubble diagram in Fig.~\ref{fig:propagators}, and we have 
\bea
H_l(p) &=& W_l(p) H(\eta)\,, \ \ \ 
H(\eta) = \psi(i\eta) + \frac{1}{2i\eta} - \ln(i\eta)\,,
\eea
where $\psi(z)$ is the digamma function. 
The nuclear interaction is represented in terms of the effective range parameters 
in the functions $K_l(p)$. Because of the modification of the counting rules~\cite{sa-prc18},
we include the effective range terms up to $p^6$ order for $l=0,1,2$ and the terms up to $p^8$ for $l=3$. We fix one of the effective range parameters by using a condition that
the denominator of the amplitudes, i.e., the inverse of the dressed $^{16}$O propagators, 
\bea
D_l(p) &=& K_l(p) - 2\kappa H_l(p)\,,
\eea
vanish at the binding energies, $E=-B_l$, where $B_l$ are the binding energies of the 
sub-threshold $0_2^+$, $1_1^-$, $2_1^+$, $3_1^-$ states of $^{16}$O with respect to the 
$\alpha$-$^{12}$C breakup state. 
Thus, $D_l(i\gamma_l)=0$ at $p=i\gamma_l$, where $\gamma_l$ are the binding momenta,
$\gamma_l=\sqrt{2\mu B_l}$, and the functions $K_l(p)$ are obtained as
\bea
K_l(p) &=& \frac12r_l(\gamma_l^2 + p^2) 
+ \frac14P_l(\gamma_l^4 - p^4) 
+ Q_l(\gamma_l^6 + p^6)
+ R_l(\gamma_l^8 - p^8)
+ 2\kappa H_l(i\gamma_l)\,,
\eea
where $R_l=0$ for $l=0,1,2$. 
The effective range parameters, $r_l$, $P_l$, $Q_l$ and $R_3$ are fitted to the experimental 
data, and one can calculate the asymptotic normalization coefficients (ANCs) of the 
bound states of $^{16}$O with respect to the $\alpha$-$^{12}$C state by using 
the Iwinski-Rosenberg-Spruch 
(IRS) formula~\cite{irs-prc84} \footnote{
The IRS formula is a standard method to deduce ANCs from 
elastic scattering data in momentum space calculations. 
Refer to, e.g., Ref.~\cite{scb-prc10}.
The higher-order terms truncated in the effective range expansion, 
as well as a difficulty in fitting the effective range parameters for $l=2$~\cite{sa-cpc25},  
would introduce systematic errors in the theoretical estimates.  
}
\bea
|C_b|_l &=& \frac{\gamma_l^l}{l!} \Gamma(l+1 +\kappa/\gamma_l)
\left(
\left|
\frac{dD_l(p)}{dp^2}
\right|_{p^2 = -\gamma_l^2}
\right)^{-1/2}\,.
\label{eq;Cbl}
\eea
We have 13 parameters for the subthreshold states of $^{16}$O,
while we fix two of them, $r_0$ and $r_2$, by using the binding energy 
of the ground $0_1^+$ state of $^{16}$O~\cite{sa-jkps18} 
and a value of the ANC of $2_1^+$ state of $^{16}$O, 
$|C_b|_2=10\times 10^4~\textrm{fm}^{-1/2}$~\cite{sa-cpc25}. 
The detailed expressions of the modifications are presented in Appendix B. 

For the scattering amplitudes for resonant states of $^{16}$O, they are presented in the Breit-Wigner like 
expression as
\bea
A_l^{(rs)} &=& - \frac{1}{p}
\frac{\frac12\Gamma_{(li)}(E)}{E - E_{R(li)} + R_{(li)}(E) + i\frac12 \Gamma_{(li)}(E)}
\eea
with 
\bea
\Gamma_{(li)}(E) &=& \Gamma_{R(li)}\frac{pC_\eta^2W_l(p)}{p_rC_{\eta_r}^2W_l(p_r)}\,,
\\
R_{(li)}(E) &=& a_{(li)}(E - E_{R(li)})^2 + b_{(li)}(E - E_{R(li)})^3 \,,
\eea
where $E_{R(li)}$ and $\Gamma_{R(li)}$ are the energy and width of resonant $l_{i\textrm{th}}^\pi$ state
of $^{16}$O\footnote{ 
The resonant energies $E_{R(li)}$ are measured from the $\alpha$-$^{12}$C breakup threshold energy of $^{16}$O, $E_{th} = 7.162$~MeV. The resonant energies measured from the ground state of $^{16}$O are obtained
by adding $E_{th}$, $E_x = E_{R(li)} + E_{th}$.} 
and $p$ is the relative momentum.
$p_r$ and $\eta_r$ are the relative momentum and the Sommerfeld factor at $E=E_{R(li)}$, 
$p_r=\sqrt{2\mu E_{R(li)}}$ and $\eta_r=\kappa/p_r$. We suppressed the $(li)$ indices on $p_r$ and $\eta_r$ for simplicity. 
$a_{(li)}$ and $b_{(li)}$ are the coefficients of the higher-order terms of the Taylor expansion 
in the denominator of the amplitudes around $E=E_{R(li)}$, which fit the shapes of the resonances. 

Using the expressions of the scattering amplitudes, one has the simple and transparent expressions of the $S$ matrices for the $l$th
partial waves as
\bea
S_l &=&
\frac{K_l(p)-2\kappa Re H_l(p) + ipC_\eta^2W_l(p)}{K_l(p)-2\kappa Re H_l(p) - ipC_\eta^2W_l(p)}
\prod_i 
\frac{E-E_{R(li)} +R_{(li)}(E) -i\frac12\Gamma_{(li)}(E)}{
E-E_{R(li)} +R_{(li)}(E) +i\frac12\Gamma_{(li)}(E)
}\,.
\eea
In Appendix C, 
the parameters of the bound and resonant states of $^{16}$O employed in the present work are summarized. 
Refer to Ref.~\cite{sa-prc23} for details. 
We display three sets of fitted values of the parameters in the tables.
The first set is the parameters fitted to the accurate phase shifts data reported
by Tischhauser {\it et al.}~\cite{tetal-prc09}. The second one is those fitted to 
the differential cross section data by employing DE, and the third one is those 
fitted to the differential cross section data by performing MCMC.

\section{Result of fitting and quantification of uncertainty}
We outline the computational framework used to extract the parameters of the cluster EFT model and to quantify their uncertainties. The approach combines global optimization and sampling algorithms 
 that are widely used in Bayesian inference and statistical data analysis.
Parameter estimation is performed by minimizing a global $\chi$-square over more than ten thousand elastic 
$\alpha$-$^{12}$C differential cross section data points spanning a broad range of energies and angles. Because the parameter space is high-dimensional and the objective highly nonlinear, we employ global rather than local search strategies. In the first stage, DE provides a robust point estimate and ensures convergence to the minimum. This solution then serves as the starting point for a MCMC analysis, which propagates statistical and systematic uncertainties to the physical observables. 

\subsection{Differential evolution}
We fit the cluster EFT model parameters by minimizing a global $\chi$-square over $N=11~392$ elastic $\alpha$-$^{12}$C differential cross section data points covering the $\alpha$-particle energies from 2.6 to 6.7 MeV and 32 laboratory angles ranging from $24.0^\circ$ to $165.9^\circ$ in the laboratory frame.
The theoretical cross sections are calculated from the partial wave amplitudes introduced in Sec.~II and transformed to the laboratory frame for direct comparison with experiment.

Before introducing the details of the fit, we note that the initial parametrization contained 41 variables in the previous works. In the $\ell=3$ channel two coefficients are found to be redundant, while in the $\ell=0$ and $\ell=2$ channels the effective range terms $r_0$ and $r_2$ are constrained by the ground state $0^+_2$ binding energy and the $2^+_1$ ANC of $^{16}$O, respectively. After applying these conditions, the number of independent parameters is reduced to 37. 
The fit therefore involves 37 free parameters: effective range parameters and resonance descriptors for partial waves up to $\ell=6$. 
All other parameters are varied within physics-motivated bounds, with resonance widths restricted to positive values. 
To enforce consistency, parameter combinations leading to 
unphysical poles near the physical sheet are penalized and 
excluded from the solution.

The fit minimizes a global $\chi$-square defined as
\begin{equation}
  \chi^2 = \sum_{k=1}^{N} \left( \frac{\sigma_k^{\mathrm{exp}} - \sigma_k^{\mathrm{th}}(\boldsymbol{\theta})}{\Delta_k} \right)^2 ,
\end{equation}
where $\sigma_k^{\mathrm{exp}}$ and $\sigma_k^{\mathrm{th}}$ are the measured and calculated cross sections, and $\Delta_k$ denotes the experimental uncertainty. $\mathbf{\theta}$ represents a set of effective range parameters.
For data subsets with quoted normalization errors we introduce floating scale factors with Gaussian penalties, which absorb correlated systematics without biasing the central values.

\begin{figure}
\centering
\includegraphics[width=1.0\linewidth]{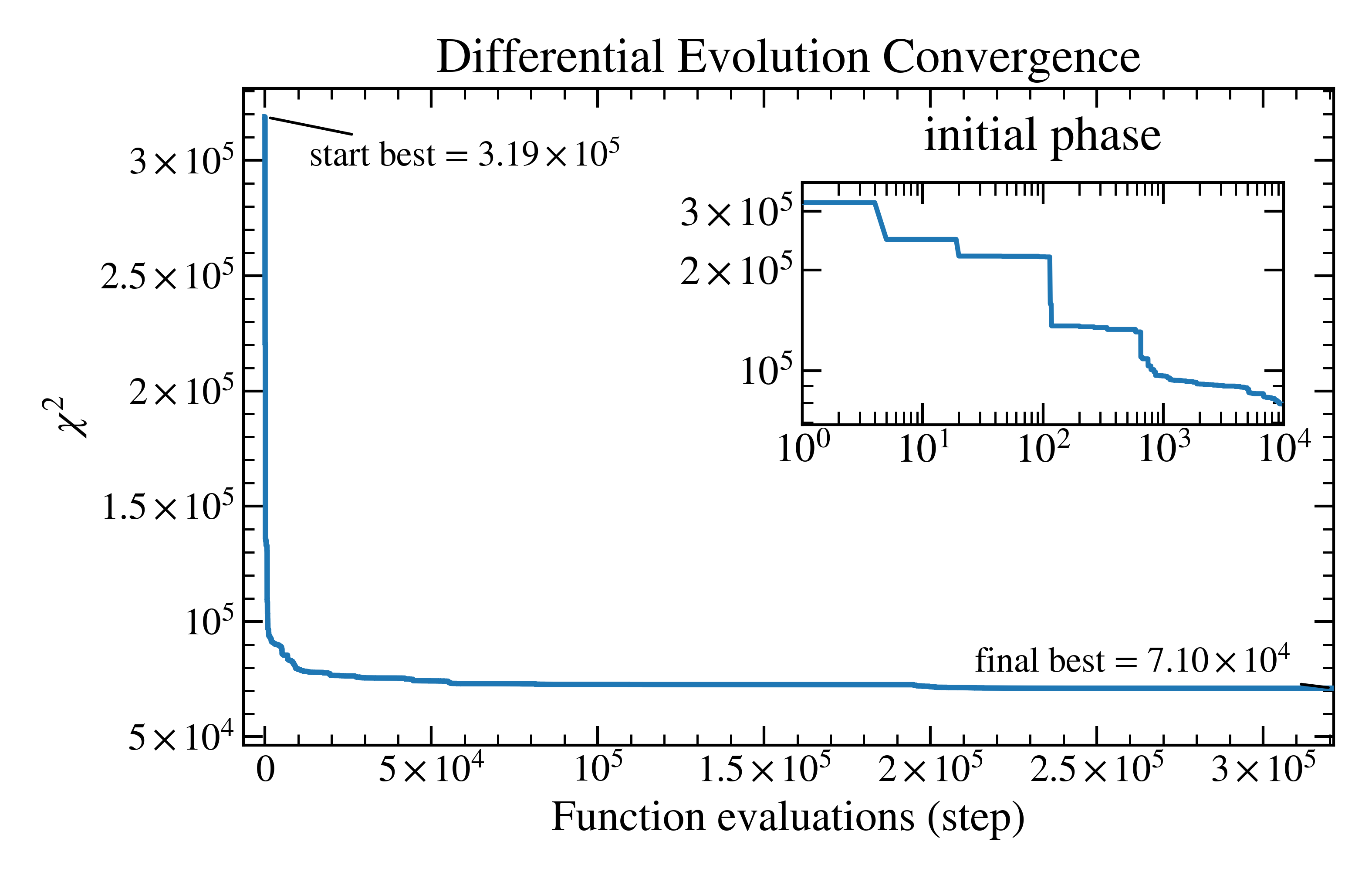}
\caption{ Differential evolution (DE) convergence for the 37-parameter fit. 
The ordinate shows the $\chi$-square, $\chi^2$, constructed from the full elastic
$\alpha$-$^{12}$C dataset ($N=11~392$ points), while the abscissa counts objective function evaluations (“step”). 
The solid curve displays the running best value,
$\min_{k\le\text{step}}\chi^2_k$, which decreases monotonically from
$\sim 3.19\times10^{5}$ at the start to $7.10\times10^{4}$ at the end of the run.
The inset (log-log scales; steps $<10^{4}$) resolves the rapid early drop and the plateau-like improvements characteristic of DE’s mutation-crossover-selection updates.
Beyond this initial phase, the main panel exhibits a slow, smooth improvement and stabilization,
consistent with robust global convergence.}
\label{fig:de-convergence}
\end{figure}

The DE algorithm is employed to perform the global optimization.
DE is a population-based, derivative-free evolutionary scheme that iteratively improves
a set of candidate solutions through differential mutation, crossover, and selection.
Trial vectors that yield a lower $\chi^2$ replace their parents, driving the population
toward the global minimum while maintaining sufficient diversity to prevent premature convergence.
This stochastic yet systematic strategy proves well suited for the present 37-parameter fit,
where the cost function exhibits multiple local minima and complex correlations among parameters.
In practice we run \texttt{scipy.optimize.differential\_evolution} with  
\texttt{strategy}=\texttt{rand2exp},
a population of 50 candidates in the 37 dimensional space, mutation factors $F\in[0.8,1.9]$, and recombination rate $CR=0.9$\,\cite{Scipy, DE-paper, DE-book}.
The initial population consists of the projected reference vector $\boldsymbol{\theta}_0$
and random draws within a prior box defined by physically motivated parameter ranges.
Each candidate is then projected back into the physically allowed domain,
ensuring that all scattering parameters satisfy the required positivity
and reality conditions imposed by the effective range formalism.
The objective combines the least-squares log-likelihood (with the optional systematic floor disabled in the final runs) and 
a mode-selectable prior; we use the \texttt{HYBRID}
setting, which adds soft penalties for grid positivity and mild anchors on $(E_R,\Gamma_R)$.
Convergence is tracked with the running-best $\chi^2$ trace, and the search terminates by a
tolerance criterion (\texttt{tol}=$10^{-6}$) or upon reaching the maximum iteration budget
(\texttt{maxiter}=$10^6$). 
For clarity, we briefly summarize the processes in one iteration of the DE
algorithm as implemented in this work. A population of candidate parameter vectors
is evolved generation by generation through mutation, crossover, and selection.
For each target vector, a mutant vector is constructed by adding weighted differences
of randomly chosen population members, and a trial vector is then formed via
crossover with a prescribed probability. The trial vector replaces its parent only
if it yields a lower value of the $\chi^2$ function. This population-based evolutionary
procedure enables an efficient exploration of the high-dimensional and non-linear
$\chi^2$ landscape of the cluster EFT, allowing the algorithm to escape local minima
and to robustly identify the global best-fit region.

The convergence behavior of the DE fit is also summarized in Fig.~\ref{fig:de-convergence}.
The plot shows the global $\chi^{2}$ value evaluated over all $N=11~392$ data points
as a function of the number of function evaluations.
An abrupt decrease occurs within the first $\sim10^{4}$ evaluations, corresponding to the
rapid elimination of high $\chi^{2}$ trial vectors during the initial exploration phase.
The inset, drawn on log-log scales, magnifies this region and reveals a sequence of plateau-like segments that reflects successive successful replacements of the population through
mutation and crossover operations.
Beyond this stage, the main panel displays a gradual and smooth improvement that stabilizes
around $\chi^{2}\!\approx\!7.1\times10^{4}$, demonstrating that the algorithm has reached a
well-defined global minimum without premature convergence.
This behavior confirms that the adopted DE configuration efficiently explores the
37-dimensional parameter space and provides a reliable set of optimized parameters
to initialize the subsequent MCMC uncertainty analysis.

\begin{figure}
\centering
\includegraphics[width=0.85\linewidth]{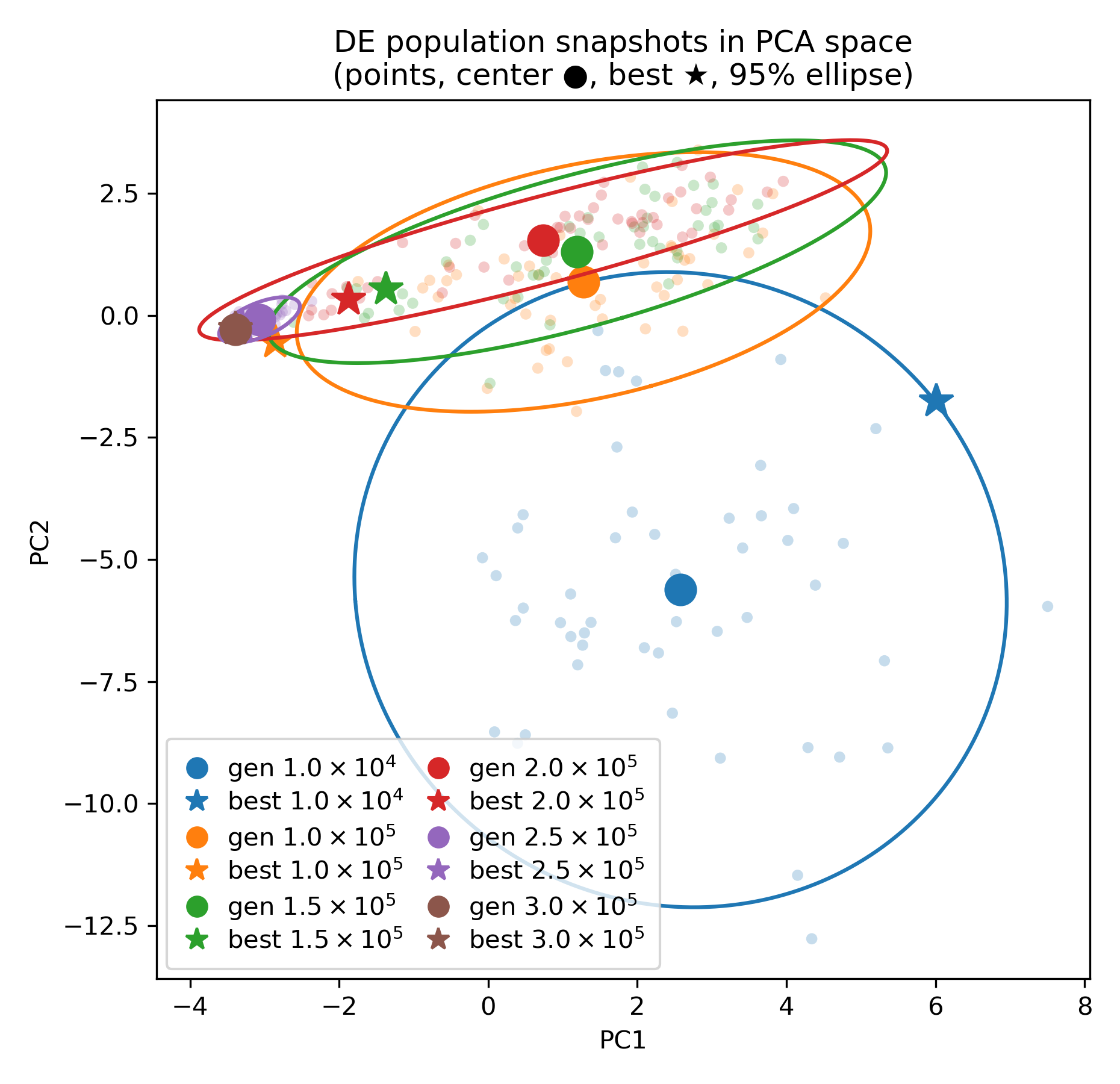}
\caption{Population snapshots of the DE optimization.
Populations from representative generations
($10^{4}$--$3\times10^{5}$) are projected onto the first two principal components
(PC1,\,PC2) of a single PCA basis fitted to the combined samples.
Points denote individual candidates, filled circles and stars mark the population centroids and best members, and ellipses show the 95\% covariance regions.
The contraction and drift of the clusters illustrate the transition from broad exploration to convergence around the global minimum.
Details of the projection and interpretation are given in the text.}
\label{fig:pca-snapshots}
\end{figure}

Figure~\ref{fig:pca-snapshots} visualizes how the DE population migrates and contracts in the
37-dimensional parameter space. Populations from generations
$10^4,\,10^5,\,1.5\times10^5,\,2\times10^5,\,2.5\times10^5,$ and $3\times10^5$
are projected onto the first two principal components (PC1,\,PC2) obtained from a single principal component analysis (PCA) fitted to the union of all shown generations; points denote individual candidates, filled circles and stars mark the population centroid and best member of each snapshot, and ellipses indicate the $95\%$ covariance regions in the projected plane.\footnote{
For visualization only, we compute a single PCA basis from the stacked population; 
PC axes are linear combinations of the original parameters and have no direct physical
meaning. The ellipses are level sets of the sample covariance in the PC plane.}
The earliest snapshot ($10^4$; blue) is widely dispersed and appears as a large ellipse away from the later clusters, reflecting broad exploration in the projected plane. As the run proceeds, the cloud drifts along a correlated direction (roughly aligned with PC1), the successive centroids move coherently, and the ellipses contract, with occasional mild re-expansion at intermediate snapshots as the population reorients upon entering a lower $\chi^2$ valley; the final snapshot ($3\times10^5$) is compact and stable.
Because PCA is a linear projection of a 37-dimensional distribution, apparent sizes and orientations reflect projected covariance and can vary somewhat with the random seed and the specific snapshots shown; the qualitative pattern is robust, however: broad initial exploration, drift along a correlated valley, and eventual contraction around a single region near the optimum.

The optimized parameter vector obtained from DE serves as the initial state for the
uncertainty analysis presented in Sec.~III~B, where MCMC propagates
the resulting parameter correlations and normalization pulls to derive credible intervals
for the cross sections, phase shifts, resonance parameters, and ANCs.
The adopted 37-parameter configuration is validated as the minimal and physically
consistent representation of the data: it simultaneously satisfies all imposed physical
constraints while achieving the lowest $\chi^2/N$ among the tested configurations.
Any further reduction of parameters violates at least one of the physical conditions or
produces a statistically significant deterioration of the fit quality.
This parametrization therefore provides a compact yet sufficient representation of the
elastic $\alpha$-$^{12}$C scattering observables within the applicable range of the cluster EFT framework.

\subsection{Markov chain Monte Carlo}
In Bayesian inference, as well as in general statistical modeling, the posterior probability distribution of the model parameters is often intractable to evaluate analytically. 
Instead of attempting to compute such complex distributions directly, the  MCMC method is widely employed to approximate the posterior distribution numerically. 
MCMC generates correlated samples that asymptotically follow the target posterior by constructing a Markov chain whose equilibrium state represents the desired distribution. 
This approach enables efficient estimation of expectation values, variances, and credible intervals of the parameters, even in high-dimensional parameter spaces.  

Compared with conventional deterministic optimization or least-squares fitting methods, which typically converge to a single solution or a local minimum, MCMC provides a global exploration of the entire parameter space through stochastic sampling. 
As a result, it allows reliable uncertainty quantification and posterior analysis even for problems that exhibit non-linear correlations, multi-modal structures, or non-Gaussian behaviors. 

In the present work, the MCMC sampling is performed using the affine-invariant ensemble sampler implemented in the \texttt{emcee} package~\cite{emcee}. 
This algorithm evolves an ensemble of multiple walkers simultaneously, enabling efficient sampling even in parameter spaces with strong anisotropy or correlations. 
The parameter space consists of 37 dimensions, 
and the ensemble comprises 200 walkers. 
Each walker is evolved for $1.2\times10^5$ steps to ensure thorough exploration of the posterior distribution, with the initial $2\times10^4$ steps discarded as burn-in to remove transient effects from initialization. 
A thinning factor of 10 is applied to reduce autocorrelation and enhance the statistical independence of the retained samples. 
Consequently, a total of approximately $2\times10^6$ independent samples are retained for posterior analysis. 
The computation is parallelized across 192 CPU cores, which significantly improves sampling efficiency and ensures robust convergence of the multidimensional posterior distribution.  

The convergence of the Markov chains is primarily assessed through the integrated autocorrelation time ($\tau_{\mathrm{int}}$). 
This quantity measures the correlation length between consecutive samples, with shorter values indicating more statistically independent draws. 
The mean autocorrelation time over all parameters is found to be 1~434.1 steps, with a minimum of 963.7 and a maximum of 2~206.8. 
Given the total number of steps per walker ($N_{\mathrm{steps}} = 1.2\times10^5$), the criterion $N_{\mathrm{steps}} \gg 50\,\tau_{\mathrm{int}}$ is well satisfied, confirming that the ensemble produces a sufficiently large number of effectively independent samples and that the Markov chains are reliably converged.

\begin{figure}[t]
\centering
\includegraphics[width=0.8\textwidth]{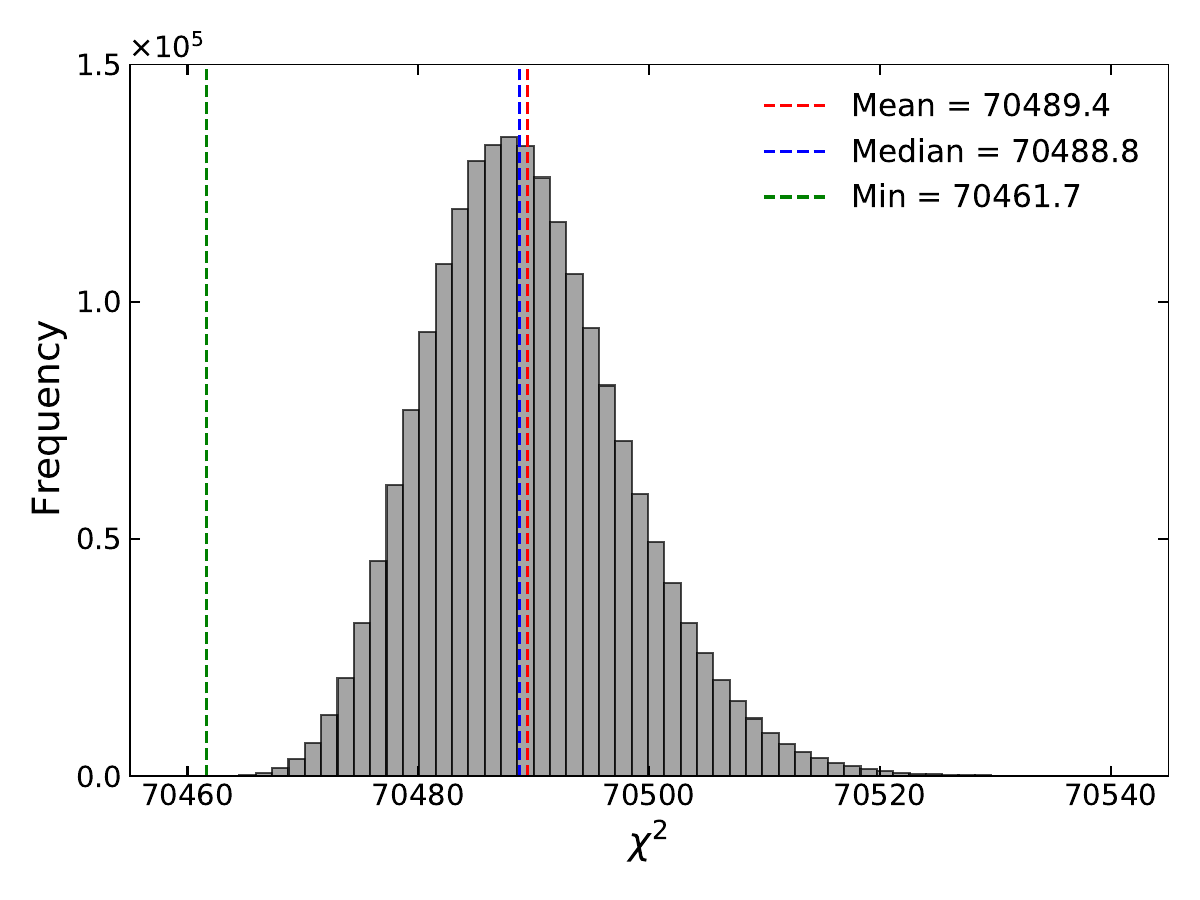}
\caption{Distribution of the total $\chi^2$ values obtained from the MCMC analysis for the elastic $\alpha$-$^{12}$C scattering.
The histogram peaks around $\chi^2 \approx 7.05\times10^4$ with mean = 70~489.4, median = 70~488.8, and minimum = 70~461.7, indicate excellent convergence and statistical stability of the parameter estimation.}
\label{fig:chi2_hist}
\end{figure}

Figure~\ref{fig:chi2_hist} shows the distribution of total $\chi^2$ values obtained from the MCMC sampling for the elastic $\alpha$-$^{12}$C scattering data.
The $\chi^2$ values are tightly clustered around $7.05\times10^4$ with a mean of 70~489.4, a median of 70~488.8, and a minimum of 70~461.7, respectively.
The narrow spread and the close agreement between the mean and the median confirm that the posterior samples are well converged, demonstrating that the model provides a stable fit to the experimental differential cross sections.
The sharply peaked $\chi^2$ distribution further demonstrates that the effective range parameters inferred from the MCMC analysis are statistically consistent and robust against sampling fluctuations.

Table~\ref{tab:chi2_summary} summarizes the statistical properties of $\chi^{2}/N$ obtained from the MCMC analysis. 
A total of $2\times10^6$ samples are used in the analysis and $N=11~392$ represents the number of experimental data points used in the fit. 
The mean value of $\chi^{2}/N = 6.1876$ with a very small standard deviation of $7.5\times10^{-4}$ indicates excellent convergence of the sampling. 
The 95\% credible interval $[6.1861,\,6.1891]$ and the best-fit value $6.1852$ demonstrate that all samples are narrowly distributed, confirming the numerical stability and robustness of the MCMC results.

\begin{table}[t]
\centering
\caption{Summary of $\chi^{2}/N$ statistics for the cross section data
with the MCMC analysis. $\sigma$ denotes the standard deviation,
and CI implies credible interval.}
\begin{tabular}{ccccc}
\hline\hline
~~~Mean ~~~& ~~~$\sigma$~~~  &~~~ Median ~~~ & ~~~95\% ($2\sigma$) CI ~~~& ~~~Best-fit~~~ \\
\hline
6.1876 & $7.5\times10^{-4}$ & 6.1875  & [6.1861, 6.1891]& 6.1852 \\
\hline\hline
\end{tabular}
\label{tab:chi2_summary}
\end{table}

\section{Result for observables}

\subsection{Differential cross section and phase shifts with uncertainties from MCMC calculation}

\begin{figure}
\centering
\includegraphics[width=1.0\linewidth]{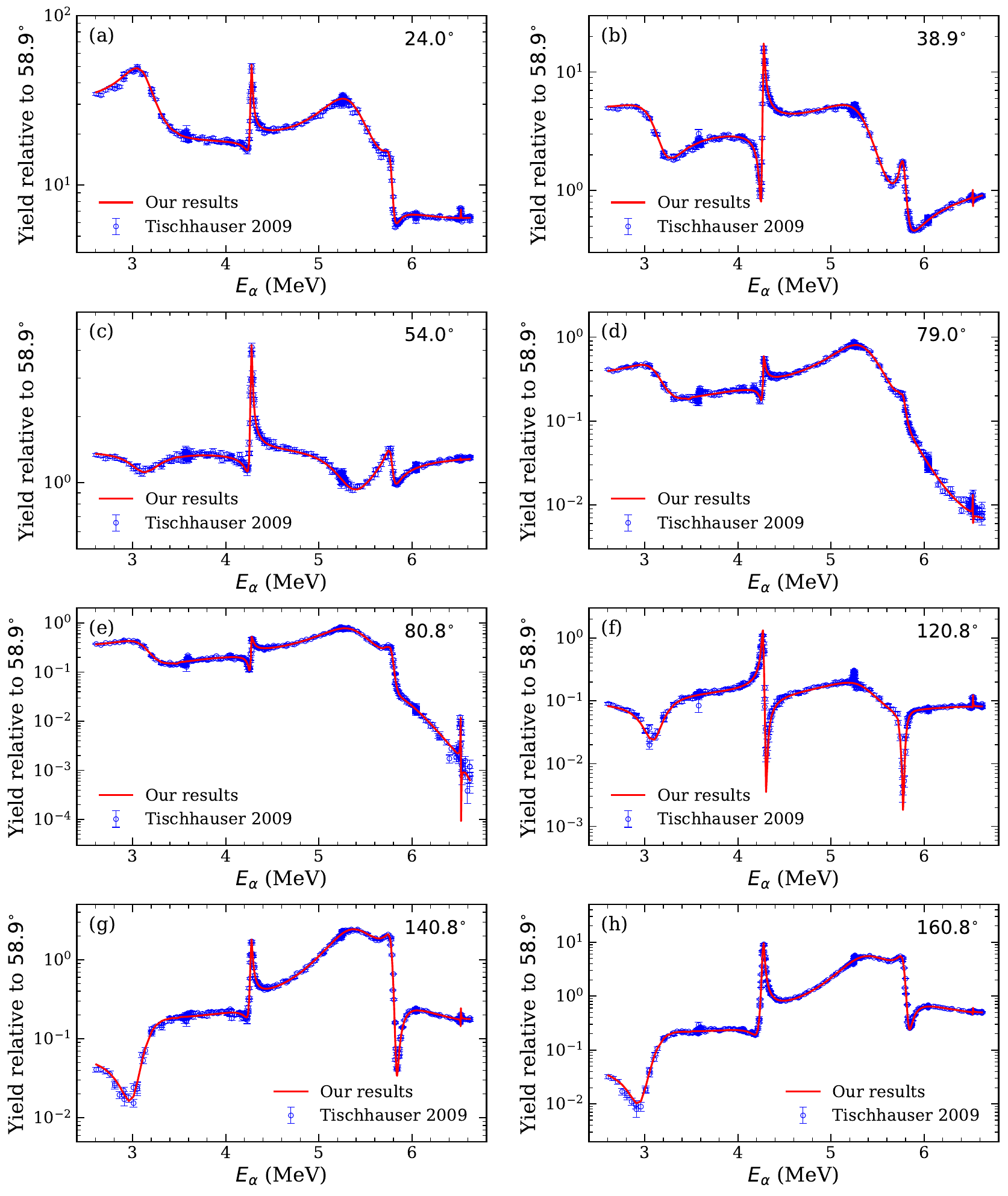}
\caption{Comparison between the calculated (red solid lines) and experimental (Tischhauser 2009~\cite{tetal-prc09}, blue circle) yields as a function of the incident energy \(E_{\alpha}\) for several scattering angles. 
The panels correspond to \(\theta_{\text{lab}} = 24.0^\circ, 38.9^\circ, 54.0^\circ, 79.0^\circ, 80.8^\circ, 120.8^\circ, 140.8^\circ,\) 
and \(160.8^\circ\), respectively. 
The yields are normalized to those at 58.9$^\circ$. 
}
\label{fig:dcs}
\end{figure}

To comprehensively assess the consistency between the experimental data and the theoretical analysis, Fig.~\ref{fig:dcs} displays the excitation curves of the elastic scattering yields for the $\alpha$-$^{12}$C system at eight representative laboratory angles ranging from $24.0^{\circ}$ to $160.8^{\circ}$ among 32 angles available from experimental data. 
The blue symbols correspond to the experimental yields reported by Tischhauser \textit{et al.}~\cite{tetal-prc09}, normalized to the yield at $58.9^{\circ}$ to emphasize the relative energy dependence across different scattering angles, and for direct comparison with results from experiment. 
The red solid curves denote the theoretical predictions obtained using the EFT with effective range parameters constrained through the MCMC fitting procedure. 
The results are presented on a logarithmic scale to clearly illustrate the variation of the yields over several orders of magnitude within the energy range $2.6~\mathrm{MeV} \leq E_{\alpha} \leq 6.7~\mathrm{MeV}$. 
As shown in Fig.~\ref{fig:dcs}, the calculated excitation curves successfully reproduce both the absolute magnitude and the energy dependence of the experimental yields across the entire angular range, indicating that the extracted EFT parameters provide a coherent and quantitatively reliable description of the dynamics in the elastic $\alpha$-$^{12}$C scattering.

\begin{figure}
\centering
\includegraphics[width=0.95\linewidth]{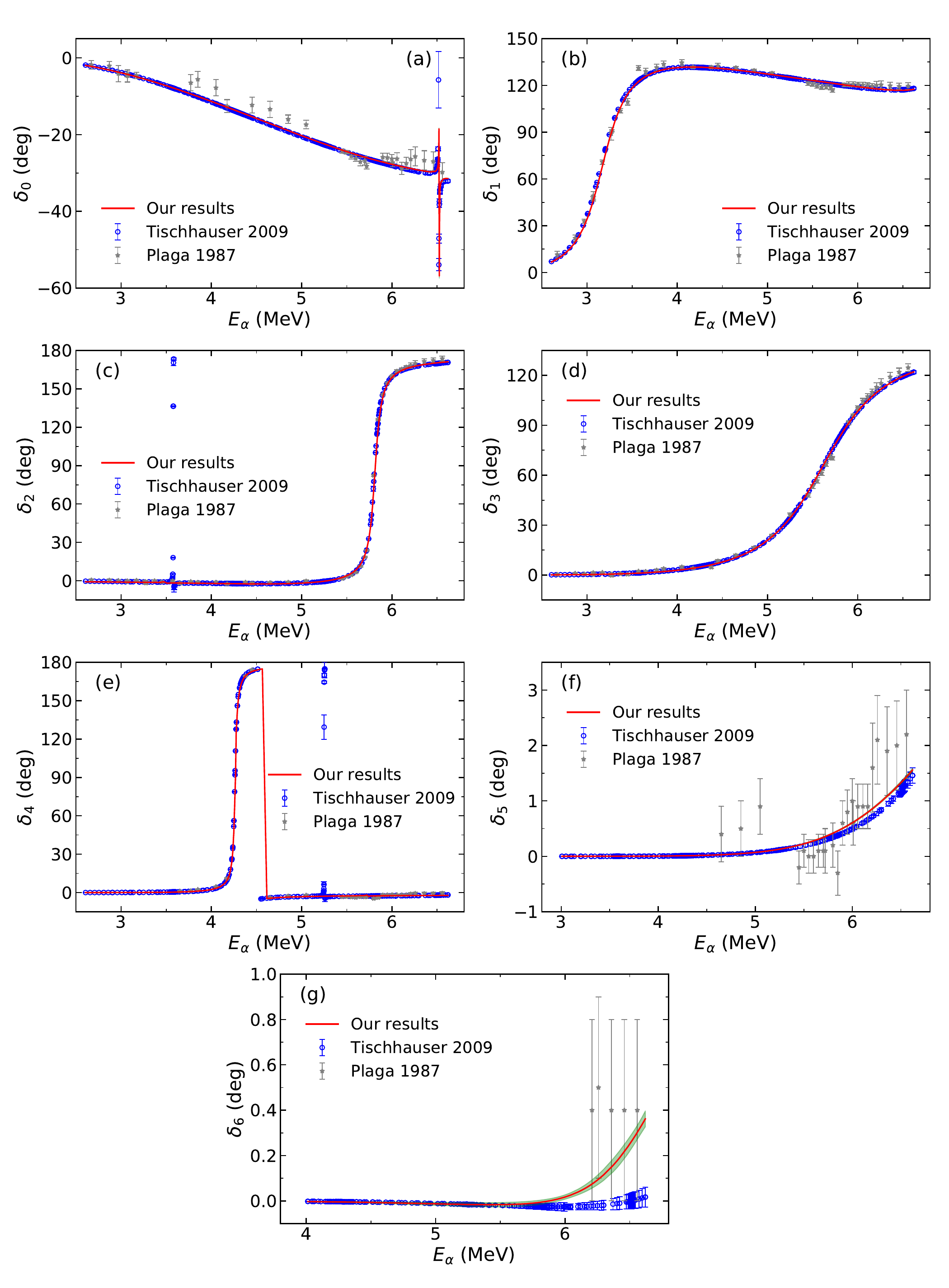}
\caption{Calculated phase shifts for $l = 0$--6 in elastic $\alpha$-$^{12}$C scattering as functions of $E_{\alpha}$ compared with $R$-matrix results~(Tischhauser 2009~\cite{tetal-prc09} and Plaga 1987~\cite{petal-npa87}).
Error ranges are denoted by green bands for each $l$.
They are unrecognizable for $l = 0$--5, but apparent for $l=6$.
}
\label{fig:PS}
\end{figure}

Figure~\ref{fig:PS} shows the calculated phase shifts $\delta_l$ ($l = 0$--6) for elastic $\alpha$-$^{12}$C scattering compared with $R$-matrix  results~\cite{tetal-prc09, petal-npa87}. 
Blue and gray points indicate the $R$-matrix results, 
while red lines represent the result from $2\times10^6$ MCMC samples. 
The model reproduces the phase shifts consistent with $R$-matrix results for low angular momenta ($l = 0, 1$) and captures the main trends for intermediate values ($l = 2$--4). 
Phase shifts for $l = 5$ and 6 are smaller than those for $l = 0$--4, and exhibit larger uncertainties but remain consistent with $R$-matrix values within the MCMC prediction range. 
Error ranges ($1\sigma$) are denoted by green bands for each $l$.
They are unrecognizable for $l = 0$--5, but apparent for $l=6$.
These results confirm that the MCMC-fitted model reliably describes elastic $\alpha$-$^{12}$C scattering across the considered angular momenta.  
Furthermore, the DE-MCMC approach captures both the non-resonant background and narrow resonant structures with widths above 20 keV without readjusting any parameters to specific input data, demonstrating the predictive power of this statistical inference method.

\begin{figure}
\centering
\includegraphics[width=0.75\linewidth]{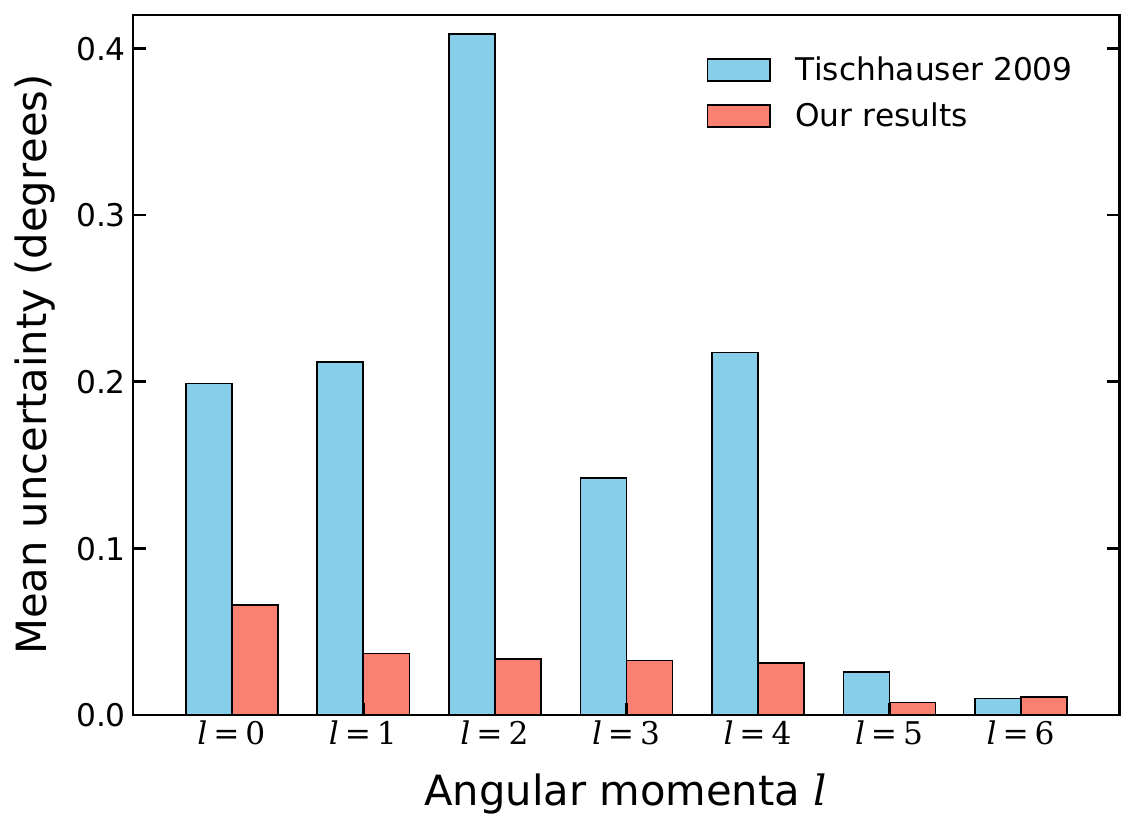}
\caption{Comparison of the average magnitudes of the error bars obtained in this work with those from Tischhauser 2009 (Ref.~\cite{tetal-prc09}). The uncertainties from the present MCMC analysis are significantly smaller than those from the previous $R$-matrix evaluation.}
\label{fig:compare_error}
\end{figure}

In Fig.~\ref{fig:compare_error}, the average magnitudes of the error bars obtained in this work are compared with those from Ref.~\cite{tetal-prc09}. 
The uncertainties extracted from the present MCMC analysis are considerably reduced compared with the previous $R$-matrix evaluation, demonstrating that the phase shifts are more precisely and stably constrained within the current framework.
Mean uncertainty of this work is largest for $l=0$, similar for $l = 1$--4, and
$l=5$, 6 exhibit similar smallest values.
While our result shows monotonic decrease of the uncertainty with increasing
$l$, $R$-matrix uncertainty exhibits strong fluctuation,
ascending behavior from $l=0$ to 3, and another
low peak structure at $l=4$ above $l=3$.
It is notable that discrepancy between $R$-matrix and this work is
biggest at $l=2$.

\subsection{Resonant energies and widths, and the ANCs of bound states of $^{16}$O}
As seen in Fig.~\ref{fig:PS}, we find the difficulty in reproducing the sharp resonant $0_3^+$, $2_2^+$, $4_2^+$ states 
of $^{16}$O in the phase shifts of $R$-matrix
for $l=0,2,4$, respectively, by using the fitted values of the parameters
to the differential cross section data. 
The sharp resonant states are clearly seen in the phase shifts reported by Tischhauser {\it et al.}~\cite{tetal-prc09}
at $E_\alpha=6.5$~MeV, 3.6~MeV, and 5.3~MeV for $l=0,2,4$, respectively. 
The fitted values of the widths in this work turn out to be 66\%, 25\%, 35\%, respectively,
compared to the values in the previous work \cite{sa-prc23}. 
The energy of the resonant $4_2^+$ state is also shifted significantly.  
The fitted values of the parameters are summarized in Appendix C. 
Thus, those sharp resonant states of $^{16}$O are less tightly constrained 
in a global fit to the differential cross section data.
We emphasize that this behavior does not indicate a deficiency of the differential
cross section data or of the $R$-matrix analysis. Rather, it reflects the limited
constraining power of a global fit for extremely narrow resonant
structures, whose contributions are confined to restricted kinematic regions and
therefore carry relatively small weight in the overall $\chi^2$. In this sense, the
properties of such sharp resonances could be more sensitive to
the fitting strategy, while the cluster EFT provides a robust description of the
global scattering behavior.
\footnote{%
Dedicated high-resolution measurements of the narrow $l=2$ and $l=4$ resonant
states in the $\alpha$-$^{12}$C system, together with refined theoretical analyses aimed
at improving their description within the cluster EFT framework, are currently in
progress~\cite{Mun2026}.
}

Regarding the other resonant states appearing in the data of Ref.~\cite{twc}, 
because the parameters of the resonant $1_2^-$ and $3_2^-$ states are included in the parameters 
of the subthreshold $1_1^-$ and $3_1^-$ states, the values of those resonant states are not fitted explicitly. 
\begin{table}
\begin{center}
\caption{
Fitted values of the energies and widths of the resonant $2_3^+$ and $4_1^+$ states of $^{16}$O. 
The values of the second column are those in the compilation edited by Tilley, Weller, and Cheves~\cite{twc}.
Those in the third column are obtained by fitting to the phase shift (PS) data \cite{sa-prc23}, 
and those in the fourth column are obtained by fitting to the differential cross section (CS) data in this work. 
}
\begin{tabular}{c|ccc}
\hline \hline
  &~~~TWC~\cite{twc}~~~&~~~~~PS~~~~~&~~~~~CS~~~~~\cr \hline
~~~$E_{R(23)}~(\textrm{MeV})$~~~&~~~~~4.358(4)~~~~~&~~~~~4.3536(1)~~~~~& ~~~~~4.35902(6)~~~~~\cr
~~~$\Gamma_{R(23)}~(\textrm{keV})$~~~&~~~~~71(3)~~~~~&~~~~~74.5(1)~~~~~&~~~~~79.1(1)~~~~~\cr
~~~$E_{R(41)}~(\textrm{MeV})$~~~&~~~~~3.194(3)~~~~~&~~~~~3.19606(1)~~~~~&~~~~~ 3.19875(2)~~~~~\cr
~~~$\Gamma_{R(41)}~(\textrm{keV})$~~~&~~~~~26(3)~~~~~&~~~~~25.91(1)~~~~~&~~~~~27.67(6)~~~~~\cr \hline\hline
\end{tabular}
\label{table:PS_CS}
\end{center}
\end{table}
In Table~\ref{table:PS_CS}, we display the fitted values of the energies and widths of the resonant $2_3^+$ and $4_1^+$ states of $^{16}$O. 
The values of the second column are those in the compilation edited by Tilley, Weller, and 
Cheves (TWC)~\cite{twc}. Those in the third and fourth columns are obtained by fitting to the phase shift data 
and the differential cross section data, respectively. 
We find a good agreement with the TWC values within the errors, except for the width of the resonant $2_3^+$ 
state. 

By using the MCMC samples of the effective range parameters, we calculate 
the values of the ANCs of the 
bound $0_1^+$, $0_2^+$, $1_1^-$, $3_1^-$ states of $^{16}$O. 
Because of the problem in fitting the ANC of the $2_1^+$ state of $^{16}$O
from the elastic $\alpha$-$^{12}$C scattering data~\cite{sa-cpc25},
we fixed the value of the ANC of the $2_1^+$ state as $|C_b|_2=10\times 10^4$~fm$^{-12}$ in the parameter fit. See Appendix B as well. 

\begin{table}
\begin{center}
\caption{The values of the ANCs 
of the bound $0_1^+$, $0_2^+$, $1_1^-$, $3_1^-$ 
states of $^{16}$O in the unit of fm$^{-1/2}$, deduced by using the effective 
range parameters fitted to the differential cross section data. 
In the previous work~\cite{sa-fbs24}, the effective range parameters are fitted to the phase shifts data 
reported by Tischhauser {\it et al.}~\cite{tetal-prc09}.}
\begin{tabular}{c|cc}
\hline \hline 
~~~~~$l_{ith}^\pi$~~~~~&~~~~~Prev. work~\cite{sa-fbs24}~~~~~&~~~~~This work~~~~~\\ \hline 
 $0_1^+$ & 45.5(3) & 46(1) \\
 $0_2^+$ & 621(9) & 660$^{+37}_{-32}$ \\
 $1_1^-$ & 1.727(3)$\times 10^{14}$ & 1.715(6)$\times 10^{14}$ \\
 $3_1^-$ & 113(8) & 75(2) \\ \hline  \hline 
\end{tabular}
\label{table:ANC}
\end{center}
\end{table}
In Table~\ref{table:ANC}, we display the deduced values of the ANCs of 
the bound states of $^{16}$O 
in the MCMC analysis with the differential cross section data,
labeled by `This work.'
Those in `Prev. work' are obtained by using the effective range
parameters fitted to the phase shift data deduced in the $R$-matrix 
analysis~\cite{tetal-prc09}.
We find that the center values of the ANC of the $0_1^+$, $0_2^+$, $1_1^-$ states
are in good agreement within the two sigma errors, while the errors of the 
present work become two to four times larger than those in the previous work. 
Because the $s$-wave parts (as well as the $p$-wave) can contribute to 
the fitting of all partial waves, the errors in this work are enhanced. 
We also find that the center value of the ANC of the $3_1^-$ state becomes smaller,
about 66\%, compared to that in the previous work, while the error is reduced
by a factor of four.

\section{Summary}
In the present work, we attempted a new methodology for the description
of elastic $\alpha$-$^{12}$C scattering.
We adopted the cluster EFT and 
expanded the scattering amplitude in terms of the effective range
parameters up to the angular momentum $l=6$.
Parameters were fitted to elastic $\alpha$-$^{12}$C cross section data
in the energy range 2.6--6.7 MeV and at angles $24.0^\circ$-$165.9^\circ$
including the resonance states below the $p$-$^{15}$N breakup threshold energy.
Using the DE algorithm,
we could find the minimum of $\chi^2$ in the parameter space globally.
With the help of DE results, MCMC analysis was facilitated.
From the analysis of correlations between parameters, we could sort out 
redundant parameters and as a result identify that the optimal number of parameters
is 37.
Accuracy of the fitting is obtained as $\chi^2/N \simeq 6.2$ in both
DE and MCMC fitting.
With the MCMC method, we also estimated the uncertainty of each parameter.

Compared with conventional fitting approaches such as gradient-based searches or manually tuned
$R$-matrix analyses, 
the global optimization framework
adopted here provides several practical and conceptual advantages.
It performs a fully automated and global exploration of the high-dimensional parameter space,
reducing the need for subjective initialization and preventing the fit from being trapped in local minima.
This allows the optimization to find a physically consistent solution even when the cost surface is highly nonlinear or multimodal.
The framework is also reproducible and easily extendable: once implemented, the same algorithmic pipeline can be applied to new datasets
or modified models with minimal human intervention, ensuring consistency across analyses.
In addition, it provides statistically rigorous uncertainty estimates through direct sampling of the posterior distribution,
so that both experimental errors and model correlations are propagated in a controlled way.
In this sense, 
the global optimization approach
does not replace traditional physics modeling but complements it,
offering an objective, scalable, and data-driven means to perform global parameter optimization
and uncertainty quantification within the cluster EFT framework.

To check the quality of the fitting, we calculated the differential cross sections 
of elastic $\alpha$-$^{12}$C scattering and compared the result to experimental data.
Our calculation reproduces the data well, and agreement to them is as good as the
$R$-matrix theory.
We also calculated the phase shifts at angular momenta from $l=0$ to
$l=6$ and estimated their uncertainties with the MCMC technique.
We find that the phase shifts agree with those of $R$-matrix calculation for 
$l \leq 4$, but substantial enhancement occurs at high energies for $l=5$ and 6
compared to the accurate estimate of $R$-matrix theory.

In the previous work~\cite{sa-prc23}, 
the phase shifts deduced in the $R$-matrix analysis
are perfectly reproduced within the cluster EFT.
However, the set of parameters that agrees with the phase shifts leads to 
a large $\chi^2$ value, $\chi^2/N \simeq 20$, for the differential cross section data. 
Thus, we fitted the parameters of the cluster EFT directly 
to the differential cross section and obtained a smaller value of $\chi^2$, 
$\chi^2/N \simeq 6.2$. 
Because the $\chi^2$ value of the differential cross section
in the $R$-matrix analysis was not mentioned in Ref.~\cite{tetal-prc09}, 
we have no measure to discuss which fit is better than the other. 
However, in the comparison of error bars in the phase shifts,
average width of error bars in each angular momentum state is notably
smaller in the present work than the $R$-matrix results in \cite{tetal-prc09}.
On the other hand, we found the difficulty in reproducing the 
sharp resonant $0_3^+$, $2_2^+$, and $4_2^+$ states of $^{16}$O.

In conclusion, we have shown that the combination of effective range expansion
in the cluster EFT and application of 
DE-MCMC optimization algorithms
to the cross section data provide a framework accurate for the description of the
elastic $\alpha$-$^{12}$C scattering process.
With an advanced fitting scheme, uncertainty could be improved compared to 
precedent works.
Advantage of the new method should be tested furthermore by applying it to
radiative capture of $\alpha$ particle by $^{12}$C and many other important
astrophysical processes.

\section*{Acknowledgements}
This work was supported by the National Research Foundation of Korea (NRF) grants funded by the Korea government (Grants No. RS-2018-NR031074, RS-2021-NR060129, RS-2022-NR070836, RS-2025-24533596, RS-2025-25400847, 2023R1A2C1003177, and RS-2025-16065411). S.-I. Ando thanks X. D. Tang for valuable discussions.

\section*{APPENDIX A: Relations between the laboratory frame 
and center-of-mass frame}

Transformation of the differential cross section and scattering angle 
between the laboratory frame and the center-of-mass frame is 
well known~\cite{i-07}. We present only the formulas necessary in the 
present work: we need two relations of the scattering angles and the 
differential cross sections in the non-relativistic limit, i.e., 
we ignored the $1/m$ corrections.
(The $\alpha$ energy, $E_\alpha$, in the laboratory frame is 
related to the energy, $E$, of the $\alpha$-$^{12}$C system in the 
center-of-mass frame as $E_\alpha=\frac43E$.)

Using the $\gamma$ factor, $\gamma=m_\alpha/m_C=1/3$, the relation of the scattering angles reads
\bea
\cos\theta &=& -\frac13\sin^2\theta_L 
\pm \frac13\sqrt{(1-\sin^2\theta_L)(9-\sin^2\theta_L)}\,,
\eea
where $\theta$ and $\theta_L$ are the scattering angles in the center-of-mass
frame and the laboratory frame, respectively. 
The relation for the differential cross sections reads
\bea
\sigma_L(\theta_L)&=& \sigma(\theta)\frac{d(\cos\theta_L)}{d(\cos\theta)}\,,
\eea
where $\sigma_L$ is the differential cross section in the laboratory frame and 
\bea
\frac{d(\cos\theta_L)}{d(\cos\theta)} &=&
\frac{1+\frac13\cos\theta}{\left(
\frac{10}{9} + \frac23\cos\theta
\right)^{3/2}}\,.
\eea

\section*{APPENDIX B: Fixing the effective range parameters, $r_0$ and $r_2$}
In this appendix, we review the methods to fix the effective range parameters, $r_0$ and $r_2$,
by using the binding energy of the ground $0_1^+$ state of $^{16}$O and a value of the ANC of 
$2_1^+$ state of $^{16}$O.
To fix $r_0$, we adopt two conditions on the inverse of the $^{16}$O propagator for $l=0$;
one is $D_0(i\gamma_{01})=0$ and the other is $D_0(i\gamma_{02})=0$, where
$\gamma_{01}$ and $\gamma_{02}$ are the binding momenta for the $0_1^+$ and $0_2^+$ states of $^{16}$O,
respectively,
$\gamma_{01}=\sqrt{2\mu B_{01}}$ and $\gamma_{02}=\sqrt{2\mu B_{02}}$.
$B_{01}$ and $B_{02}$ are the binding energies of $0_1^+$ and $0_2^+$ states of $^{16}$O 
with respect to the $\alpha$-$^{12}$C state. 
By using the two conditions mentioned above, one has~\cite{sa-prc24,sa-prc25}
\bea
D_0(p) &=&  - \frac14\left[
\gamma_{01}^2\gamma_{02}^2 
+ (\gamma_{01}^2 + \gamma_{02}^2) p^2
+ p^4
\right] P_0 
\nnb \\ && 
+ \left[
-\gamma_{01}^4\gamma_{02}^2
-\gamma_{01}^2\gamma_{02}^4
- (\gamma_{01}^4 + \gamma_{01}^2\gamma_{02}^2 + \gamma_{02}^4)p^2
+ p^6 
\right] Q_0
\nnb \\ &&
-2\kappa\left[
\frac{\gamma_{02}^2 + p^2}{\gamma_{01}^2 - \gamma_{02}^2} H_0(i\gamma_{01})
-\frac{\gamma_{01}^2 + p^2}{\gamma_{01}^2 - \gamma_{02}^2} H_0(i\gamma_{02})
+ H_0(p)
\right] \,.
\label{eq;D0}
\eea

To fix $r_2$, one may use Eq.~(\ref{eq;Cbl}). Thus, one has
\bea
r_2 &=& - \frac12 \gamma_2^4 \left[ \frac{\Gamma(3 + \kappa/\gamma_2)}{|C_b|_2} \right]^2
            + \frac{1}{12}\kappa^3
            \nnb \\ && 
            - \left(P_2 + \frac{17}{20}\kappa \right) \gamma_2^2
                   -6 \left(Q_2 - \frac{757}{4032\kappa}\right)\gamma_2^4
            - \frac{289}{1260\kappa^3}\gamma_2^6 
            + \frac{2455}{11088\kappa^5} \gamma_2^8\,.
\eea

\section*{APPENDIX C: Effective range parameters}
The numerical values of the effective range parameters are summarized in Tables IV--VI. The parameters for $l = 0$ and 1 are listed in Table IV, those for $l = 2$ and 3 in Table V, and those for $l = 4-6$ in Table VI. See the text for details.
\begin{table}[t]
\caption{Fitted parameters for $l = 0$ and $l = 1$.
Three values are presented for each parameter.
The values in the first row are obtained from fits to the phase-shift data,
while those in the second and third rows are determined by fitting the differential cross-section data using the DE method and MCMC analysis, respectively.
Parameters marked with an asterisk ($*$) are fixed by other experimental data,
and blank entries indicate that the corresponding parameters are not included in the fit.}
\begin{tabular}{c|ccccc}
\hline \hline
$l^\pi_{ith}$   & $p^0$      & $p^2$      & $p^4$                 & $p^6$                  \cr \hline 
$0_1^+,0_2^+$   & $a_0$~(fm) & $r_0$~(fm) & $P_0~(\textrm{fm}^3)$ & $Q_0~(\textrm{fm}^5)$  \\
Phase shift     & *          & *          & $-0.03452(2)$         & 0.001723(7)            \cr
DE              & *          & *          & $-0.034365$           & 0.001677               \cr 
MCMC            & *          & *          & $ -0.034423(58)$      & $0.001688(20)$         \cr \hline
$0_3^+$         & $E_{R(03)}~(\textrm{MeV})$ & $\Gamma_{R(03)}~(\textrm{keV})$ & &  \cr
                & 4.8883(1)                  & 1.35(3)                         & &  \cr 
                & 4.892853                   & 0.917                           & &  \cr 
                & 4.892851(15)               & 0.891(11)                       & &  \cr \hline
$0_4^+$ & $E_{R(04)}~(\textrm{MeV})$ & $\Gamma_{R(04)}$ & $a_{(04)}~(\textrm{MeV}^{-1})$ & $b_{(04)}~(\textrm{MeV}^{-2})$  \cr 
        & *                          & *                & 0.756(7)                       & 0.167(4)                        \cr 
        & *                          & *                & 0.843                          & 0.217                           \cr 
        & *                          & *                & $0.815^{+0.017}_{-0.010}$      & $0.195^{+0.009}_{-0.008}$       \cr \hline \hline
$1_1^-, 1_2^-$ & $a_1~(\textrm{fm}^3)$ & $r_1~(\textrm{fm})$ & $P_1~(\textrm{fm}^{-1})$ & $Q_1~(\textrm{fm}^{-3})$  \cr 
               & *                     & 0.415314(7)         & $-0.57428(7)$            & 0.02032(2)                \cr
               & *                     & 0.415386            & $-0.57383$               & 0.02038                   \cr
               & *                     & 0.415342(12)        & $-0.57430(12)$           & 0.02023(4)                \cr \hline
$1_3^-$ & $E_{R(13)}~(MeV)$ & $\Gamma_{R(13)}~(\textrm{keV})$ & $a_{(13)}~(\textrm{MeV}^{-1})$ & $b_{(13)}~(\textrm{MeV}^{-2})$  \cr
        & *                 & *                               & 0.43(25)                       & 3.8(7)                          \cr 
        & *                 & *                               & 0.22                           & 7.6                             \cr  
        & *                 & *                               & $0.28^{+0.10}_{-0.11}$         & 7.7(5)                          \cr \hline \hline
\end{tabular}
\label{table:fitted_parameters-1}
\end{table}

\begin{table}[h]
\caption{Fitted parameters for $l = 2$ and $ l= 3$. 
See the caption of Table~\ref{table:fitted_parameters-1} for detailed descriptions.}
\begin{tabular}{c|ccccc}
\hline \hline
$l^\pi_{ith}$   & $p^0$                 & $p^2$                    & $p^4$                    & $p^6$               & $p^8$  \cr \hline 
$2_1^+$         & $a_2~(\textrm{fm}^3)$ & $r_2~(\textrm{fm}^{-3})$ & $P_2~(\textrm{fm}^{-1})$ & $Q_2~(\textrm{fm})$ & \cr 
                & *                     & *                        & $-1.049(2)$              & 0.141(2)            & \cr
                & *                     & *                        & $-1.054$                 & 0.130               & \cr 
                & *                     & *                        & $-1.092(2)$              & 0.111(1)            & \cr \hline
$2_2^+$ & $E_{R(22)}~(\textrm{MeV})$       & $\Gamma_{R(22)}~(\textrm{keV})$ & & & \cr
        & 2.68308(1)                       & 0.76(1)                         & & & \cr
        & 2.61925                          & 0.73                            & & & \cr
        & $2.61191^{+0.00227}_{-0.00207}$  & 0.19(6)                         & & & \cr \hline
$2_3^+$ & $E_{R(23)}~(\textrm{MeV})$ & $\Gamma_{R(23)}~(\textrm{keV})$ & $a_{(23)}~(\textrm{MeV}^{-1})$ & $b_{(23)}~(\textrm{keV})$ & \cr
        & 4.3536(1)                  & 74.5(1)                         & 1.2(1)                         & 0.6(1)                    & \cr
        & 4.35897                    & 79.3                            & 1.19                           & 1.50                      & \cr 
        & 4.35902(6)                 & 79.1(1)                         & 0.96(4)                        & 1.17(5)                   & \cr \hline
$2_4^+$ & $E_{R(24)}~(\textrm{MeV})$ & $\Gamma_{R(24)}~(\textrm{keV})$ & $a_{(24)}$ & $b_{(24)}$ &  \cr
        & 5.89(2) & 237(19) & 0.67(9) & 0.2(1) &  \cr
        & *       & *       &         &        &  \cr
        & *       & *       &         &        &  \cr \hline \hline
$3_1^-$,$3_2^-$ & $a_3~(\textrm{fm}^7)$ & $r_3~(\textrm{fm}^{-5})$ & $P_3~(\textrm{fm}^{-3})$ & $Q_3~(\textrm{fm}^{-1})$ & $R_3~(\textrm{fm})$ \cr
                & *                     & 0.0355(2)                & $-0.446(9)$              & 0.311(5)                 & $-0.152(3)$ \cr
                & *                     & 0.0343                   & $-0.436$                 & 0.313                    & $-0.152$    \cr
                & *                     & 0.0361(3)                & $-0.389(8)$              & 0.336(4)                 & $-0.138(2)$ \cr \hline
$3_3^-$ & $E_{R(33)}~(\textrm{MeV})$ & $\Gamma_{(R(33)}~(\textrm{keV})$ & $a_{(33)}~(\textrm{MeV}^{-1})$ & $b_{(33)}~(\textrm{MeV}^{-2})$ & \cr
        & *                          &  *                               & 32(32)                         & $3.2(32)\times 10^2$           & \cr
        & *                          &  *                               &                                &                                & \cr
        & *                          &  *                               &                                &                                & \cr \hline\hline
\end{tabular}
\label{table:fitted_parameters-2}
\end{table}

\begin{table}[h]
\caption{Fitted parameters for $l = 4$, $l = 5$, and $l = 6$. 
See the caption of Table~\ref{table:fitted_parameters-1} for detailed descriptions.}
\begin{tabular}{c|ccccc}
\hline \hline
$l^\pi_{ith}$ & $p^0$                      & $p^2$                           & $p^4$                          & $p^6$      \cr \hline 
$4_1^+$       & $E_{R(41)}~(\textrm{MeV})$ & $\Gamma_{R(41)}~(\textrm{keV})$ & $a_{(41)}~(\textrm{MeV}^{-1})$ & $b_{(41)}~(\textrm{MeV}^{-2})$ \cr
              & 3.19606(1)                 & 25.91(1)                        & 0.740(3)                       & 0.304(5)  \cr
              & 3.19878                    & 27.71                           & 0.741                          & 0.608     \cr
              & 3.19875(2)                 & 27.67(6)                        & 0.387(34)                      & 1.028(37) \cr \hline
$4_2^+$ & $E_{R(42)}~(\textrm{MeV})$ & $\Gamma_{R(42)}~(\textrm{keV})$ & &  \cr
        & 3.93655(2)                 & 0.425(4)                        & &  \cr
        & 4.29785                    & 0.340                           & &  \cr
        & 4.29617(123)               & 0.147(28)                       & &  \cr\hline
$4_3^+$ & $E_{R(43)}~(\textrm{MeV})$ & $\Gamma_{R(43)}~(\textrm{keV})$ & $a_{(43)}~(\textrm{MeV}^{-1})$ & $b_{(43)}~(\textrm{MeV}^{-2})$  \cr
        & *                          & *                               & 0.889(6)                       & 0.216(3)  \cr
        & *                          & *                               & 0.421                          & 0.039     \cr
        & *                          & *                               & 0.411(5)                       & 0.036(1)  \cr\hline\hline
$5_1^-$ & $E_{R(51)}~(\textrm{MeV})$ & $\Gamma_{R(51)}~(\textrm{keV})$ & $a_{(51)}~(\textrm{MeV}^{-1})$ & $b_{(51)}~(\textrm{MeV}^{-2})$  \cr
        & *                          & *                               & 0.572(6)                       & 0.104(2)  \cr
        & *                          & *                               & 0.602                          & 0.114     \cr
        & *                          & *                               & 0.580(12)                      & 0.102(4)  \cr\hline\hline
$(bg)$ & & $r_6~(\textrm{fm}^{-11})$ & $P_6~(\textrm{fm}^{-9})$ &  \cr
       & & $-0.3(2)$                 & 2(1)                     &  \cr
       & & $-0.20$                   & 0.22                     &  \cr
       & & $-0.24^{+0.35}_{-0.38}$       & $1.96^{+1.48}_{-1.10}$   &   \cr \hline \hline
$6_1^+$ & $E_{R(61)}~(\textrm{MeV})$ & $\Gamma_{R(61)}~(\textrm{keV})$ & $a_{(61)}~(\textrm{MeV}^{-1})$ & $b_{(61)}~(\textrm{MeV}^{-2})$  \cr
        & *                          & *                               & 0.8(1)                         & 0.18(4)  \cr
        & *                          & *                               & 0.723                          & 0.134    \cr
        & *                          & *                               & $0.751^{+0.029}_{-0.024}$      & $0.146^{+0.029}_{-0.024}$   \cr \hline\hline
\end{tabular}
\label{table:fitted_parameters-3}
\end{table}
 
\end{document}